、
\documentclass[sigconf]{acmart}
\usepackage{bbding}
\usepackage{multirow}
\usepackage{array}
\usepackage{enumitem}
\usepackage{xspace}
\usepackage{balance}
\newcommand{\tool}{{BackdoorMBTI}\xspace}
\newcommand{\ie}{i.e.,\xspace}
\newcommand{\eg}{e.g.,\xspace}

\AtBeginDocument{%
  }

\copyrightyear{2025}
\acmYear{2025}
\setcopyright{cc}\setcctype{BY}
\acmConference[KDD '25]{Proceedings of the 31st ACM SIGKDD Conference on
Knowledge Discovery and Data Mining V.1}{August 3--7, 2025}{Toronto, ON,
Canada}
\acmBooktitle{Proceedings of the 31st ACM SIGKDD Conference on Knowledge
Discovery and Data Mining V.1 (KDD '25), August 3--7, 2025, Toronto, ON,
Canada}
\acmDOI{10.1145/3690624.3709385}
\acmISBN{979-8-4007-1245-6/25/08}

\begin{document}

\title{BackdoorMBTI: A Backdoor Learning Multimodal Benchmark Tool Kit for Backdoor Defense Evaluation}

\author{Haiyang Yu}
\authornote{Authors contributed equally to this research.}
\orcid{0009-0008-7942-7125}
\affiliation{%
  \institution{Shanghai Jiao Tong University}
  \streetaddress{800 Dong Chuan Road}
  \city{Shanghai}
  \country{China}
  \postcode{200240}
}
\email{haiyang_yu@sjtu.edu.cn}
\author{Tian Xie}
\authornotemark[1]
\orcid{0009-0009-5421-3064}
\affiliation{%
  \institution{Shanghai Jiao Tong University}
  \streetaddress{800 Dong Chuan Road}
  \city{Shanghai}
  \country{China}
  \postcode{200240}
} 
\email{xietian1164567053@sjtu.edu.cn}
\author{Jiaping Gui}
\authornotemark[1]
\authornote{Corresponding authors.}
\orcid{0009-0001-4272-9604}
\affiliation{%
  \institution{Shanghai Jiao Tong University}
  \streetaddress{800 Dong Chuan Road}
  \city{Shanghai}
  \country{China}
  \postcode{200240}
}
\email{jgui@sjtu.edu.cn}
\author{Pengyang Wang}
\orcid{0000-0003-3961-5523}
\affiliation{%
  \institution{University of Macau}
  \streetaddress{Avenida da Universidade}
  \city{Taipa}
  \state{Macau SAR}
  \country{China}
  \postcode{999078}
}
\email{pywang@um.edu.mo}
\author{Pengzhou Cheng}
\orcid{0000-0003-0945-8790}
\affiliation{%
  \institution{Shanghai Jiao Tong University}
  \streetaddress{800 Dong Chuan Road}
  \city{Shanghai}
  \country{China}
  \postcode{200240}
} 
\email{cpztsm520@sjtu.edu.cn}
\author{Ping Yi}
\authornotemark[2]
\orcid{0000-0003-4530-5118}
\affiliation{%
  \institution{Shanghai Jiao Tong University}
  \streetaddress{800 Dong Chuan Road}
  \city{Shanghai}
  \country{China}
  \postcode{200240}
} 
\email{yiping@sjtu.edu.cn}
\author{Yue Wu}
\authornotemark[2]
\orcid{0000-0002-6107-7859}
\affiliation{%
  \institution{Shanghai Jiao Tong University}
  \streetaddress{800 Dong Chuan Road}
  \city{Shanghai}
  \country{China}
  \postcode{200240}
} 
\email{wuyue@sjtu.edu.cn}

\renewcommand{\shortauthors}{Haiyang Yu et al.}


\begin{abstract}
Over the past few years, the emergence of backdoor attacks has presented significant challenges to deep learning systems, allowing attackers to insert backdoors into neural networks. When data with a trigger is processed by a backdoor model, it can lead to mispredictions targeted by attackers, whereas normal data yields regular results. The scope of backdoor attacks is expanding beyond computer vision and encroaching into areas such as natural language processing and speech recognition. Nevertheless, existing backdoor defense methods are typically tailored to specific data modalities, restricting their application in multimodal contexts. While multimodal learning proves highly applicable in facial recognition, sentiment analysis, action recognition, visual question answering, the security of these models remains a crucial concern. Specifically, there are no existing backdoor benchmarks targeting multimodal applications or related tasks. 

In order to facilitate the research in multimodal backdoor, we introduce \tool, the first backdoor learning toolkit and benchmark designed for multimodal evaluation across three representative modalities from eleven commonly used datasets. \tool provides a systematic backdoor learning pipeline, encompassing data processing, data poisoning, backdoor training, and evaluation. The generated poison datasets and backdoor models enable detailed evaluation of backdoor defenses. Given the diversity of modalities, \tool facilitates systematic evaluation across different data types. Furthermore, \tool offers a standardized approach to handling practical factors in backdoor learning, such as issues related to data quality and erroneous labels. We anticipate that \tool will expedite future research in backdoor defense methods within a multimodal context. Code is available at \url{https://github.com/SJTUHaiyangYu/BackdoorMBTI}.
\end{abstract}




\maketitle

\section{Introduction}

With the advancement and widespread use of artificial intelligence (AI), neural networks have become an integral component of our modern life, handling diverse data from various devices and applications. However, they face growing threats from backdoor attacks, which are rapidly evolving and present real-world risks. Users may encounter poison data, where attackers infiltrate specific triggers into datasets before training, potentially impacting all users of these compromised datasets. As neural networks scale up, training costs also increase, forcing users to rely on third-party training resources that may lack security, thereby highlighting the real threat posed by backdoor attacks in practical scenarios. To counter the impact of backdoor attacks, researchers have proposed various countermeasures, including backdoor detection methods~\cite{ac, neuroninspect, ma2019nic, deepinspect}, model repair techniques~\cite{anp, clp, finepruning, li2021neural}, and so on. In recent years, several backdoor learning benchmarks~\cite{backdoorbench, backdoorbox, trojanzoo, bagdasaryan2021blind} have also been developed. However, they follow the same path as the defense work and mainly focus on one specific domain.

Modern backdoor attacks are not confined to images and have broadened their scope to include text and audio-related scenarios in the real world~\cite{li2022survey}. Some attacks even support multiple modalities simultaneously~\cite{hammoud2023look, han2024backdooring}. However, current efforts are primarily focused on designing efficient algorithms within specific domains such as computer vision. For example, highly integrated benchmarks like TrojanZoo are difficult to extend to new modalities. Although BackdoorBench supports both images and text, it separates the implementations of these two modalities due to differences in data processing, model loading, and training. These existing benchmarks are practically inadequate, highlighting an urgent need for modality support. 

The aforementioned problem stems from multiple factors. Firstly, the complexity of real-world data makes it challenging to conduct research and evaluate defense methods effectively in multimodal areas, resulting in uncertainties about their effectiveness. Secondly, the absence of a standardized evaluation baseline makes it hard to provide an objective evaluation of different algorithms. Lastly, current solutions often overlook empirical factors such as noise, undermining their overall efficacy in the real world. Nonetheless, addressing these issues is challenging. This is driven by the inherent complexity of the migration task, particularly within the framework of a multimodal application that necessitates support for diverse modalities and models. Additionally, attacks and defenses may involve numerous private parameters in their settings, making it difficult to evaluate them under a standard baseline.

To address the above issues, we introduce \tool, a novel and unified benchmark and toolkit, dedicated to the evaluation of multimodal backdoor learning. Our benchmark comprises three key aspects: 1) \tool supports three modalities and incorporates eleven representative datasets, seventeen attacks, and seven defense methods. It encompasses diverse classification task scenarios and extends attacks beyond computer vision to include audio and text domains. Furthermore, we adapt the defense methods to a multimodal context and release an open-source backdoor learning benchmark for image, text, and audio backdoor learning. 2) \tool provides easy-to-access backdoor datasets and establishes a standard benchmark for evaluating backdoor defenses in multimodal settings. 3) \tool takes into account factors such as low-quality data and erroneous labels, which align with real-world scenarios.


\tool offers a unified pipeline that ensures fair evaluation within a multimodal context, distinguishing it from other benchmarks. While existing backdoor learning benchmarks primarily focus on unimodal tasks, especially in computer vision, multimodal backdoor learning remains largely unexplored despite the ubiquity of diverse modal data in real-world applications. Table~\ref{table_modality_support} illustrates the modality support in current benchmarks. We can see that only BackdoorBench~\cite{backdoorbench} and Backdoor101~\cite{bagdasaryan2021blind} support both image and text. However, the two modalities implemented in these two benchmarks are integrated separately. To the best of our knowledge, we are the first to design a benchmark encompassing all three (\ie image, text, and audio) backdoor learning modalities.

In our experiments, \tool has demonstrated the ability to effectively handle the aforementioned issues. 
We have also incorporated a noise generator to simulate noises from both data and labels in the real world. Our experimental findings suggest that, while noises may not significantly impact the performance of backdoor attacks, they play a crucial role in backdoor defenses.


We summarize our contributions as follows:
\begin{itemize}
	\item \tool integrated eleven datasets, seventeen attacks, and seven defenses, covering diverse application scenarios, such as object classification, facial recognition, sentiment analysis, speech command recognition, and speaker identification. 
	\item To enhance reproducibility and facilitate multimodal backdoor research, \tool offers an open-source backdoor learning framework that supports image, text, and audio tasks. We have developed a unified evaluation pipeline that is both user-friendly and easily extensible. Additionally, \tool provides the community with accessible poisoned datasets and models for defense evaluation. The source codes, accompanied by user-friendly documents, are accessible at \url{https://github.com/SJTUHaiyangYu/BackdoorMBTI}.
	
	\item To improve practicality, we include noise factors as a robustness assessment module, simulating real-world backdoor defense applications with low-quality data and erroneous labels. Our findings, from a multimodal and empirical perspective, indicate that noise factors enhance model robustness, thereby improving defense performance.	
\end{itemize}

\begin{table}[]
\caption{The modality support in current benchmarks.}
\label{table_modality_support}
\begin{tabular}{llll}
\toprule
Benchmarks    & Image        & Text         & Audio        \\ 
\midrule
TrojAI \cite{trojai}        & \Checkmark   &  & \\
TrojanZoo \cite{trojanzoo}    & \Checkmark   & & \\
BackdoorBench \cite{backdoorbench} & \Checkmark   & \Checkmark   & \\
BackdoorBox \cite{backdoorbox}   & \Checkmark   & & \\
OpenBackdoor \cite{openbackdoor}  & & \Checkmark   & \\
Backdoor101 \cite{bagdasaryan2021blind} & \Checkmark   & \Checkmark   & \\
Ours          & \Checkmark   & \Checkmark   & \Checkmark   \\ 
\bottomrule
\end{tabular}
\end{table}

\vspace{-\baselineskip}
\section{Related Work}

\begin{table*}[ht]
\caption{An overview of 11 datasets included in \tool. *CelebA is a face attributes classification dataset, and all the face attributes are labeled in a bin format, in order to do the multi-class classification task on it, we follow the same settings in \cite{wang2022bppattack}.}
\label{tab_ds}
\centering
\begin{tabular}[\textwidth]{lllllp{4.7cm}}
\toprule
Task                                  & Dataset                 & Modality& Classes & Total Instances & Used In Experiment \\
\midrule
\multirow{5}{*}{Object Classification}& CIFAR10 \cite{cifar10}  & Image & 10      & 60,000          & \cite{turner2018clean, shafahi2018poison, saha2020hidden, turner2019label, nguyen2020input, wang2022bppattack, zeng2021rethinking, doan2021lira, nguyen2021wanet, souri2022sleeper, li2021neural, anp, tran2018spectral, li2021anti, huang2022backdoor, clp, i-bau, chen2022effective, zheng2022pre, zhao2020bridging, abs, qi2023towards, wang2022rethinking, mntd, tao2022model, wang2022training, tang2021demon, guo2023scale, zhu2023selective, sha2022fine, li2021backdoor, spectre} \\
                                      & TinyImageNet \cite{tiny, imagenet} & Image & 200     & 110,000         & \cite{lv2023data, shafahi2018poison, saha2020hidden, liu2020reflection, bagdasaryan2021blind, doan2021lira, li2021invisible, souri2022sleeper, li2021anti, huang2022backdoor, clp, chen2022effective, zheng2022pre, abs, qi2023towards, wang2022rethinking, guo2019tabor} \\
\midrule
\multirow{2}{*}{Traffic Sign Recognition}              & GTSRB \cite{gtsrb} & Image & 43      & 51,839          & \cite{lv2023data, barni2019new, yao2019latent, liu2020reflection, nguyen2020input, wang2022bppattack, zeng2021rethinking, doan2021lira, nguyen2021wanet, li2021neural, wang2019neural, li2021anti, i-bau, abs, qi2023towards, deepinspect, wang2022rethinking, tao2022model, neuroninspect, wang2022training, tang2021demon, zhu2023selective, sha2022fine, strip, guo2019tabor} \\
\midrule
Facial Recognition                    & CelebA \cite{celeba} & Image & 8*       & 202,599         & \cite{lv2023data, wang2022bppattack, nguyen2021wanet} \\
\midrule
\multirow{2}{*}{Sentiment Analysis}   & SST-2 \cite{sst2} & Text  & 2       & 11,855          & \cite{openbackdoor, qi2020onion} \\
                                      & IMDb \cite{imdb} & Text  & 2       & 50,000          & \cite{lv2023data, bagdasaryan2021blind, strip, yang2021rap, openbackdoor, bki} \\
\midrule
\multirow{2}{*}{Topic Classification} & DBpedia \cite{dbpedia} & Text  & 14      & 630,000         & \cite{bki} \\
                                      & AG's News \cite{agnews} & Text  & 4       & 31,900          & \cite{qi2020onion, openbackdoor} \\
\midrule
Speech Command Recognition            & SpeechCommands \cite{speechcommands} & Audio & 35      & 105,829         & \cite{mntd, finepruning, strip} \\
\midrule
Music Genre Classification            & GTZAN \cite{gtzan} & Audio & 10      & 1,000           & - \\
\midrule
Speaker Identification                & VoxCeleb1 \cite{voxceleb1} & Audio & 1251    & 100,000         & - \\
\bottomrule
\end{tabular}
\end{table*}


\subsection{Backdoor Attack}

Existing backdoor attacks are typically categorized into three types, data poisoning attacks, training control attacks, and model modification attacks~\cite{guo2022overview, li2022survey}. Data poisoning attacks involve the adversary manipulating the training data only \cite{gu2017badnets, chen2017targeted, nguyen2021wanet, liu2020reflection, li2021invisible, liu2018trojaning, barni2019new, souri2022sleeper, zeng2021rethinking}, while training control attacks allow the adversary to not only manipulate the data but also control the training process~\cite{yao2019latent, doan2021lira, bagdasaryan2021blind, nguyen2020input, li2021backdoorinph}. Model modification attacks, as described in~\cite{bai2022hardly, tang2020embarrassingly, qi2022towards}, enable the adversary to manipulate the model directly. 

Backdoor attacks can also be categorized into various other types. For example, from the perspective of attack goals, adversaries aim to achieve higher stealthiness~\cite{turner2019label, chen2017targeted, nguyen2021wanet, liu2020reflection, li2021invisible, barni2019new, souri2022sleeper, doan2021lira}, pursue a higher attack success rate~\cite{souri2022sleeper, li2021invisible}, or launch attacks without modifying labels~\cite{turner2019label, liu2020reflection, li2021invisible, barni2019new, souri2022sleeper}. To enhance stealthiness, researchers are focusing on designing invisible patches that are imperceptible to human eyes when applied to image data. Regarding label modification, it can be divided into clean-label attacks, where the label remains unchanged, and dirty-label attacks. Additionally, there are other types of attacks, such as those with sample-specific dynamic patterns~\cite{nguyen2021wanet, li2021invisible, doan2021lira}, where each data sample has a unique corresponding backdoor patch.

Backdoor attack research has primarily focused on computer vision. However, in recent years, this research direction has broadened its scope, extending beyond images to include text and audio~\cite{openbackdoor, audiosurvey, hammoud2023look}. To support these emerging types of backdoor attacks, we design \tool to work in a multimodal paradigm.

\subsection{Backdoor Defense}

Backdoor defense in multi-modal scenarios pertains to operations (\eg detection, mitigation, or removal) targeting backdoor triggers across various modalities (\eg images, text, audio). The existing defense methods can be categorized according to the machine learning lifecycle as follows:

\textbf{Preprocessing.} The goal of defenders is to destroy backdoor patterns on instances. This is motivated by the observation that backdoor attacks may lose their efficacy when the trigger used during the attack differs from the one used during the poisoning process, ShrinkPad \cite{li2021backdoor} follows the paradigm to prevent backdoor triggering. In the study by \cite{liu2017neural}, an autoencoder is employed as a preprocessor defined between the input and the neural network to address this issue. Additionally, in Deepsweep \cite{qiu2021deepsweep}, researchers explored 71 data augmentation techniques to counter backdoor attacks. 

\textbf{Poison suppression.} The goal of defenders is to prevent backdoor injections during the training process. For instance, ABL \cite{li2021anti} accomplishes poisoning suppression through a two-step approach, which is backdoor sample identification and unlearning. In DBD \cite{huang2022backdoor}, Huang et al. prevent the clustering observed in backdoor injection via self-supervised learning and semi-supervised fine-tuning. NONE \cite{wang2022training} proposed a training approach to prevent the generation of hyperplane for backdoors based on the observation backdoor neurons create a hyperplane in the affected labels through piecewise linear functions.

\textbf{Backdoor removing and mitigation.} The objective of the defenders is to eliminate or reduce the impact of backdoor effects in backdoor models, typically achieved through fine-tuning and pruning. In \cite{finepruning}, the effectiveness of fine-tuning (FT) and fine-pruning (FP) in defending against backdoor attacks is demonstrated. Consequently, significant research has focused on how to identify backdoor neurons, which is crucial for pruning-based methods. ANP \cite{anp} identifies backdoor neurons using adversarial perturbations, CLP \cite{clp} employs channel-based Lipschitz sensitivity, DBR \cite{chen2022effective} uses the sensitivity metric feature consistency towards transformations, while I-BAU \cite{i-bau} formulates it as a minimax problem to unlearn triggers. MCR \cite{zhao2020bridging} enhances model robustness using mode connectivity, SEAM \cite{zhu2023selective} utilizes catastrophic forgetting to erase backdoor triggers, SFT \cite{sha2022fine} demonstrates the effectiveness of fine-tuning in eliminating most advanced backdoors, and NAD \cite{li2021neural} employs attention distillation.

\textbf{Backdoor detection.} There are two primary categories for detecting backdoor attacks in machine learning models: data-level detection and model-level detection. The main goal of model-level detection is to determine if a model contains a backdoor, as referenced in \cite{abs, deepinspect, mntd, wang2019neural, neuroninspect, chou2020sentinet, strip, guo2019tabor}. Similarly, the main goal of data-level detection is to determine if a dataset contains poisoned samples, in data-level method, \cite{ac, liu2017neural, zeng2021rethinking, wang2019neural, udeshi2022model, qi2020onion, yang2021rap, tran2018spectral, guo2023scale, tang2021demon} are input filtering methods, which detects poison samples in reference stage.  

\textbf{Trigger reverse.} In trigger reverse methods, defenders aim to identify potential backdoor triggers. The process of trigger reverse involves searching for a specific input pattern within the model that can serve as a trigger. The identification of such a trigger regards the model as a backdoor model; otherwise, it is considered a benign model. Importantly, it could find natural backdoor triggers compared with other methods. For instance, NC \cite{wang2019neural} provides an optimization method to find possible triggers. FeatureRE \cite{wang2022rethinking} introduces a utilization of feature space constraints to uncover backdoor triggers, leveraging the observation that both input-space and feature-space backdoors are associated with feature space hyperplanes.

In our benchmark, we aim to encompass all the mentioned categories. Currently, at least one method has been implemented in each category, except for preprocessing, which will be supported in the future.

\subsection{Backdoor Benchmark}

Several benchmarks have been proposed in the field of backdoor attacks, such as TrojAI \cite{trojai}, TrojanZoo \cite{trojanzoo}, BackdoorBench \cite{backdoorbench}, BackdoorBox \cite{backdoorbox}, and Backdoor101 \cite{bagdasaryan2021blind}. TrojAI is a closed platform primarily focused on evaluating model detection defenses and is mainly utilized for backdoor model detection competitions. TrojanZoo is an end-to-end benchmark that includes attacks, defenses, and evaluations. BackdoorBench offers a multitude of experiments on image datasets, including 8,000 trials, and provides analyses and visualizations. BackdoorBox integrates backdoor attacks and defenses, with flexible invocation methods, making it a user-friendly framework for backdoor learning. Backdoor101 adds support for federated learning.

Compared with the above benchmarks, \tool distinguishes itself in several key aspects: 1) \tool supports three types of data, \ie image, text, and audio. While existing benchmarks can only support the first two, as shown in Table~\ref{table_modality_support}. 2) Users can directly access the backdoor poison dataset generated in our framework. 3) \tool considers real-world factors, which enables the generation of low-quality and erroneous label data.
\section{Supported Datasets and Models}
\begin{figure*}[!htbp]
  \includegraphics[width=\textwidth]{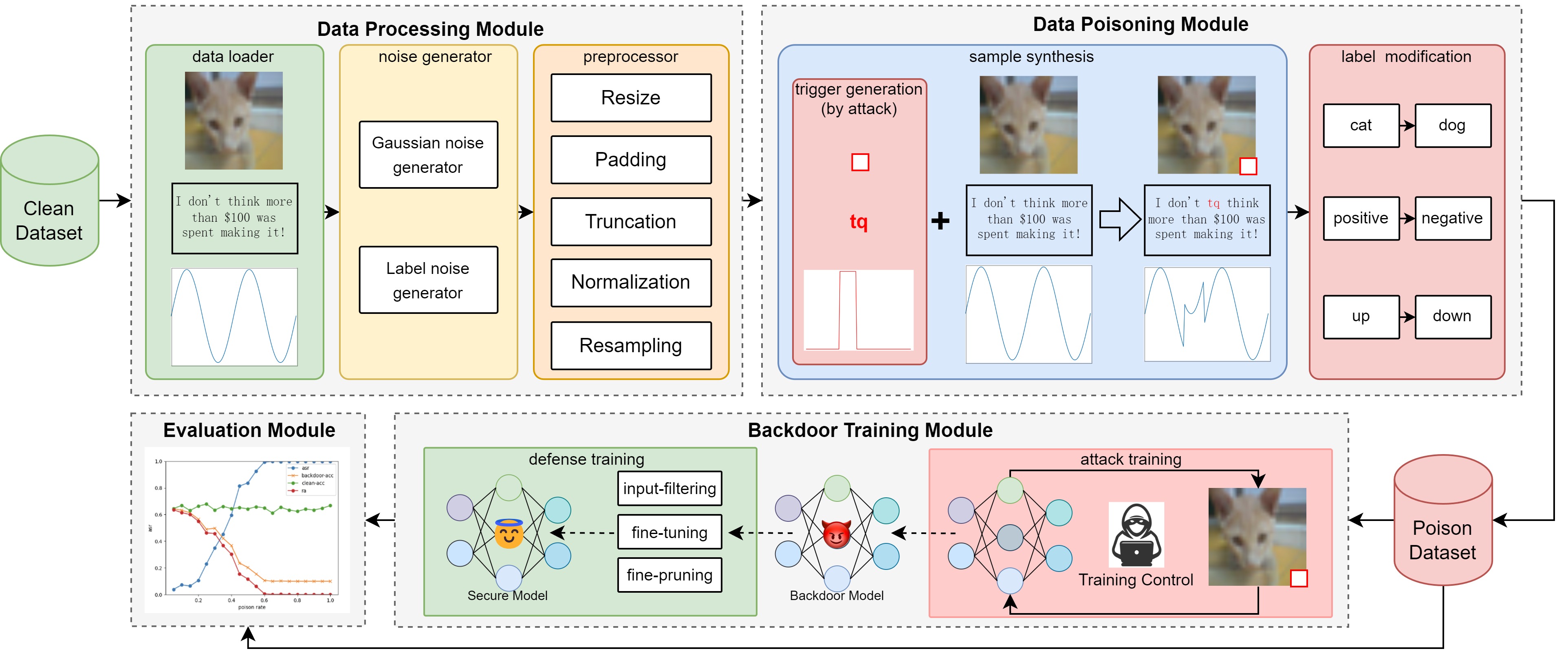}
  \caption{The architecture overview of \tool.}
  \Description{pipeline of backdoor poisoned dataset generation}
  \label{fig:arch}
\end{figure*}

The \tool framework includes 11 datasets, as shown in Table \ref{tab_ds}, covering 8 different tasks. One significant reason for selecting these 11 datasets is their public availability, ensuring easy access. Additionally, these datasets are widely used in current backdoor learning research experiments. For each modality, we have selected a commonly used model for experiments; for instance, ResNet for computer vision, BERT for natural language processing, and CNN for speech recognition. A total of 22 models have been implemented in our benchmark (refer to our repository page), with the flexibility to easily extend it with additional models.

\section{Implemented Attack and Defense Algorithms}

There are 17 backdoor attacks in different modalities and 7 backdoor defenses implemented in our framework.
In this section, we provide a brief overview of the implemented attack and defense algorithms.

\subsection{Implemented Attack Algorithms}
\begin{table}[!htbp]
\centering
\caption{The implemented attacks in \tool. Sample specific (SS), whether the trigger is sample specific.}
\label{tab_atk}
\begin{tabular}{cllllll}
\toprule
\multicolumn{1}{l}{Modality}
                        & Attack & Visible & Pattern & Add & SS \\
\midrule
\multirow{8}{*}{Image} & BadNets \cite{gu2017badnets} & Visible & Local & Yes & No \\
                        & BPP \cite{wang2022bppattack} & Invisible & Global & Yes & No \\
                        & SSBA \cite{li2021invisible} & Invisible & Global & No & Yes \\
                        & WaNet \cite{nguyen2021wanet} & Invisible & Global & No & Yes \\
                        & LC \cite{turner2019label} & Invisible & Global & No & Yes \\
                        & SBAT \cite{sbat} & Invisible & Global & No & Yes \\
                        & PNoise \cite{pnoise} & Invisible & Global & Yes & Yes \\
                        & DynaTrigger \cite{dynatrigger} & Visible & Local & Yes & Yes \\
\midrule
\multirow{5}{*}{Text} & BadNets \cite{gu2017badnets} & Visible & Local & Yes & No \\
                        & AddSent \cite{addsent} & Visible & Local & Yes & No \\
                        & SYNBKD \cite{synbkd}& Invisible & Global & No & Yes \\
                        & LWP \cite{lwp} & Visible & Local & Yes & No \\
                        & BITE\cite{yan2022bite} & Invisible & Local & Yes & Yes \\
\midrule
\multirow{4}{*}{Audio} & Blend \cite{chen2017targeted} & - & Local & Yes & No \\
                        & DABA \cite{liu2022opportunistic} & - & Global & Yes & No \\
                        & GIS \cite{koffas2023going} & - & Global & No & No \\
                        & UltraSonic \cite{koffas2022can} & - & Local & Yes & No \\
\bottomrule
\end{tabular}
\end{table}

We have integrated 17 backdoor attacks into our multimodal benchmark, with some adapted from the computer vision domain for text and audio applications, such as BadNets and Blend. Others were originally proposed within their respective domains.

As depicted in Table \ref{tab_atk}, we have chosen BadNets, LC and Blend attacks to represent classic backdoor attacks, while WaNet, BPP, SBAT, PNoise and DynaTrigger represent the latest backdoor attacks in the vision domain. In addition to character-based backdoor attacks like BadNets in text, our benchmark also supports sentence-level backdoor attacks, such as AddSent \cite{addsent} and SYNBKD \cite{synbkd}. For audio data, we have adapted Blend attacks and provided an implementation for DABA \cite{liu2022opportunistic}, GIS \cite{koffas2023going} and UltraSonic \cite{koffas2022can} attacks.

\subsection{Implemented Defense Algorithms}

\begin{table}[!htbp]
\caption{The implemented defenses in \tool. a) Input: BM, Backdoor Model; CD, Clean Dataset; PD, Poison Dataset, a dataset including malicious backdoor data in it. b) Stage: IT, In Training stage; PT, Post Training stage. c) Output: CM, Clean Model, backdoor model after migration or a secure-training model; CD, Clean Dataset, a sanitized dataset; TP, Trigger Pattern, backdoor trigger pattern reversed by the defense.}
\label{tab_def}
\begin{tabular}{ccccccccc}
\toprule
\multirow{2}{*}{Defense} & \multicolumn{3}{c}{Input} & \multicolumn{2}{c}{Stage} &  \multicolumn{3}{c}{Output} \\ \cline{2-9} 
                                 & BM & CD & PD & IT & PT & CM & CD & TP\\
\midrule
STRIP \cite{strip} & \checkmark & \checkmark & & & \checkmark &  & \checkmark  &\\
AC \cite{ac} & \checkmark & & \checkmark & & \checkmark & \checkmark & \checkmark &\\
FT \cite{finepruning} & \checkmark & \checkmark & & \checkmark & & \checkmark & &\\
FP \cite{finepruning} & \checkmark & \checkmark & & & \checkmark & \checkmark & &\\
ABL \cite{li2021anti} & \checkmark & & \checkmark & \checkmark & & \checkmark & & \\
CLP \cite{clp} & \checkmark & & & \checkmark & & \checkmark & & \\
NC \cite{wang2019neural} & \checkmark & \checkmark&  & & \checkmark & \checkmark & & \checkmark\\
\bottomrule
\end{tabular}
\end{table}

In our multimodal framework, we have implemented seven different backdoor defense methods. When selecting these defense methods, we first considered their theoretical applicability to multimodal scenarios from three perspectives: 1) general principle: for instance, STRIP uses entropy, which is applicable across data types. 2) common techniques, such as tuning and pruning, adapt across modalities. 3) practical observations, for example, CLP leverages minor perturbations (as backdoor models are sensitive to them) for detection. We also took into account the categories of these defense methods and chose recent works as references.

As indicated in Table \ref{tab_def}, STRIP \cite{strip} is a data-level sample detection method that focuses on input filtering, while ABL \cite{li2021anti} is a poison suppression method designed to prevent backdoor insertion during the training phase. FT \cite{finepruning}, FP~\cite{finepruning}, and CLP \cite{clp} are post-training methods that can be easily adapted for other data types as they are type-independent. Lastly, NC \cite{wang2019neural} serves as the representative for trigger reverse methods.

\section{Architecture}

To standardize the evaluation on multimodal and to facilitate future research in multimodal backdoor learning, we have developed a multimodal backdoor learning toolkit. A key difference between \tool and existing benchmarks on backdoor learning is its consideration of real-world noise factors, \ie low-quality data and erroneous labels. In this section, we will outline the framework architecture and discuss the design of two crucial components in our framework: the noise generator and the backdoor poisoner.

\subsection{Architecture Overview}
Figure \ref{fig:arch} depicts the architectural overview of \tool. The framework covers the entire pipeline of backdoor learning in a multimodal context, including four key modules: data processing, data poisoning, backdoor training, and evaluation. We explain each of the modules in more detail below.

1) Data Processing. The data processing module comprises three primary components: the data loader, noise generator, and preprocessor. Within this module, the clean dataset is loaded using the data loader, the noise generator is applied to each data item, and the preprocessor outputs a standardized data item. In contrast to other benchmarks that only support one data type, we have implemented various preprocessing techniques to support multimodal data. This includes resizing and normalization for images, tokenization and word embedding for text, and audio resampling. To simulate real-world applications, we introduce Gaussian noise as data noise and mislabeling to emulate natural label noise.

2) Data Poisoning. The data poisoning module processes standardized data and generates poisoned data as output. The major component of this module is the backdoor poisoner, which is responsible for executing data poisoning tasks. These tasks include trigger generation, synthesizing poisoned samples, and modifying labels. Each attack included in our benchmark has its unique trigger generation process, which is also integrated into the backdoor poisoner.

3) Backdoor Training. The backdoor training module is responsible for the training task using generated datasets and models. It consists of two distinct training pipelines: one for attack training and the other for defense training. The backdoor attack training pipeline imitates the standard training procedure but replaces the training dataset with the poison dataset. In the defense pipeline, the backdoor model created during attack training is utilized, and defense methods are applied either during training or after training. Since our backdoor poisoner is implemented as a dataset class wrapper, it can slow down training speed as the GPU waits for trigger generation and sample synthesis after fetching data from the disk. To mitigate this issue, we separate backdoor poisoning and training processes, generating the backdoor poison dataset before training in our pipeline. Additionally, for training control backdoor attacks, which typically follow a distinct training process, we implement their training procedure separately. We integrate this type of attack into the common backdoor training module (as shown in Figure~\ref{fig:arch}) to establish a standardized training pipeline.

4) Evaluation. The evaluation module takes a backdoor model and a curated test set as input and outputs performance metrics. The curated test set is specifically designed for evaluation purposes. It is generated using the data poisoning module with a poison ratio of 100\%, wherein all instances with the attack target label are excluded. Detailed information about attack and defense performance metrics can be found in Table~\ref{tab:results}.

\begin{table*}[ht]
\caption{The performance overview of backdoor attacks and defenses. CIFAR-10, SST-2, and SpeechCommands are used in the experiments for image, text, and audio modality separately. Metrics used include Clean Accuracy (CAC) for clean model classification, Backdoor Accuracy (BAC) for backdoor model classification, Attack Success Rate (ASR) for successful triggered attacks, Robustness Accuracy (RAC) for correct classification despite trigger presence, and Detection Accuracy (DAC), recall (REC), and F1 score for evaluating backdoor detection methods.}
\resizebox{\textwidth}{!}{
\begin{tabular}{llcccccccccccccccccccccccc}
\toprule
\multirow{2}{*}{Model} & Defense→ & \multicolumn{3}{c}{No Defense} & \multicolumn{3}{c}{AC} & \multicolumn{3}{c}{STRIP} & \multicolumn{3}{c}{ABL} & \multicolumn{3}{c}{FT} & \multicolumn{3}{c}{FP}& \multicolumn{3}{c}{CLP}& \multicolumn{3}{c}{NC} \\
                & Attack↓ & CAC & ASR & RAC & DAC & REC & F1 & DAC & REC & F1 & BAC & ASR & RAC& BAC & ASR & RAC & BAC & ASR & RAC  & BAC & ASR& RAC & BAC & ASR & RAC\\
\midrule
Resnet18 & BadNets-mislabel &77.31 & 94.93 & 4.31 & 63.57 & 33.58 & 14.97 & 86.62 & 84.22 & 54.58 & 60.65 & 1.07 & 64.20 & 78.00 & 3.58 & 75.47 & 76.96 & 0.33 & 30.47 & 52.91 & 20.34 & 43.50 & 48.33 & 0.13 & 15.65\\
Resnet18 & BadNets-noise &75.13 & 94.87 & 4.26 & 53.49 & 45.11 & 15.62 & 75.59 & 68.57 & 34.91 & 59.28 & 0.00 & 66.19 & 76.78 & 2.41 & 74.83 & 71.80 & 1.33 & 38.53 & 70.99 & 49.17 & 36.56 & 51.35 & 0.51 & 37.78\\
Resnet18 & BadNets-normal &77.10 & 94.86 & 4.29 & 53.96 & 45.11 & 15.62 & 88.95 & 54.97 & 48.73 & 54.11 & 0.00 & 61.16 & 78.35 & 3.49 & 75.41 & 70.16 & 2.37 & 24.23 & 39.55 & 13.37 & 30.62 & 48.91 & 0.06 & 19.02\\
Resnet18 & BPP-mislabel &77.29 & 83.27 & 12.90 & 67.38 & 29.35 & 14.66 & 84.21 & 71.25 & 46.29 & 40.45 & 0.10 & 36.28 & 78.37 & 5.73 & 52.91 & 77.26 & 8.36 & 35.64 & 36.57 & 18.63 & 18.77 & 50.24 & 0.04 & 18.89\\
Resnet18 & BPP-noise &74.94 & 82.33 & 13.20 & 56.14 & 43.05 & 15.78 & 80.80 & 84.10 & 45.54 & 56.27 & 0.16 & 47.20 & 77.11 & 7.54 & 50.07 & 54.83 & 3.33 & 25.02 & 50.11 & 20.12 & 24.22 & 46.50 & 7.02 & 22.34\\
Resnet18 & BPP-normal &77.30 & 83.22 & 12.78 & 54.26 & 43.79 & 15.45 & 83.55 & 63.23 & 42.33 & 24.68 & 0.00 & 22.21 & 78.68 & 7.02 & 52.90 & 67.51 & 12.08 & 33.32 & 60.37 & 42.92 & 24.88&/ &/ & / \\
Resnet18 & SSBA-mislabel &77.85 & 99.96 & 0.03 & 66.39 & 29.60 & 14.39 & 86.93 & 11.23 & 14.09 & 60.74 & 0.27 & 66.54 & 78.19 & 2.71 & 76.24 & 77.12 & 0.03 & 39.12 & 77.61 & 7.14 & 71.71 & 49.23 & 1.94 & 25.35\\
Resnet18 & SSBA-noise &75.36 & 99.67 & 0.26 & 55.33 & 43.96 & 15.81 & 77.62 & 19.25 & 14.11 & 57.69 & 0.02 & 65.47 & 77.07 & 2.16 & 75.46 & 71.74 & 8.56 & 37.28 & 67.69 & 12.51 & 60.01 & 49.77 & 1.48 & 28.54\\
Resnet18 & SSBA-normal &77.90 & 99.92 & 0.08 & 53.75 & 45.78 & 15.89 & 86.23 & 9.70 & 11.86 & 58.85 & 0.01 & 66.86 & 79.27 & 2.68 & 76.56 & 27.14 & 41.08 & 15.07 & 76.74 & 2.50 & 75.54 & 48.12 & 6.46 & 16.89\\
Resnet18 & WaNet-mislabel &77.85 & 99.52 & 0.36 & 68.40 & 27.13 & 14.08 & 77.62 & 99.18 & 45.83 & 26.91 & 15.09 & 49.53 & 78.78 & 6.51 & 39.38 & 77.05 & 0.13 & 16.88 & 77.99 & 99.53 & 0.34 & 48.30 & 2.80 & 31.24\\
Resnet18 & WaNet-noise &75.31 & 99.22 & 0.57 & 55.27 & 44.06 & 15.83 & 73.28 & 99.90 & 41.65 & 28.45 & 0.69 & 60.93 & 77.06 & 62.61 & 13.40 & 65.90 & 2.98 & 20.19 & 12.75 & 0.00 & 20.93 & 49.54 & 3.54 & 41.93\\
Resnet18 & WaNet-normal &77.96 & 99.37 & 0.48 & 53.96 & 44.37 & 15.54 & 80.89 & 98.85 & 49.69 & 26.53 & 1.68 & 57.19 & 78.33 & 9.00 & 48.77 & 70.61 & 1.99 & 19.43 & 23.05 & 23.17 & 13.14 & 50.52 & 1.89 & 32.61\\
\midrule
CNN & Blend-mislabel &88.77 & 97.04 & 2.64 & 64.24 & 31.54 & 14.37 & 69.03 & 25.89 & 13.73 & 82.03 & 96.00 & 3.34 & 90.47 & 0.15 & 85.84 & 69.62 & 9.67 & 52.49 & 88.76 & 97.04 & 2.63&/ &/ &/\\
CNN & Blend-noise &87.99 & 95.72 & 3.88 & 63.71 & 32.37 & 14.51 & 58.76 & 35.82 & 14.18 & 27.81 & 0.02 & 24.08 & 89.33 & 0.29 & 83.15 & 74.44 & 3.25 & 46.39 & 87.98 & 95.72 & 3.88&/ &/ &/\\
CNN & Blend-normal &89.64 & 95.64 & 3.92 & 64.27 & 31.83 & 14.49 & 61.16 & 37.75 & 15.61 & 81.12 & 93.52 & 5.32 & 91.02 & 0.26 & 84.58 & 71.93 & 1.26 & 53.14 & 89.64 & 95.64 & 3.92&/ &/ &/\\
CNN & DABA-mislabel &88.16 & 92.04 & 6.50 & 64.31 & 31.38 & 14.33 & 37.45 & 68.41 & 17.23 & 80.85 & 91.75 & 5.94 & 90.58 & 15.22 & 8.48 & 63.31 & 15.33 & 6.99 & 88.15 & 92.03 & 6.40&/ &/ &/\\
CNN & DABA-noise &87.53 & 91.92 & 6.49 & 64.10 & 32.27 & 14.61 & 63.09 & 23.23 & 15.21 & 33.32 & 17.37 & 4.68 & 89.48 & 2.05 & 8.39 & 62.97 & 13.39 & 6.20 & 87.53 & 91.92 & 6.48&/ &/ &/\\
CNN & DABA-normal &88.58 & 91.85 & 6.65 & 64.27 & 31.99 & 14.56 & 50.40 & 46.77 & 15.21 & 4.42 & 25.81 & 2.40 & 90.70 & 31.44 & 7.72 & 50.60 & 0.71 & 6.79 & 88.57 & 91.84 & 6.65&/ &/ &/\\
CNN & GIS-mislabel &87.94 & 97.31 & 0.43 & 64.26 & 31.65 & 14.42 & 63.12 & 35.11 & 15.34 & 75.27 & 98.80 & 0.19 & 90.88 & 1.43 & 16.23 & 50.17 & 41.21 & 5.68 & 87.94 & 97.31 & 0.43&/ &/ &/\\
CNN & GIS-noise &87.16 & 93.65 & 1.22 & 64.21 & 31.93 & 14.51 & 62.90 & 33.15 & 14.53 & 43.18 & 65.96 & 6.39 & 88.99 & 0.98 & 14.40 & 44.30 & 49.22 & 4.92 & 87.16 & 93.65 & 1.21&/ &/ &/\\
CNN & GIS-normal &89.67 & 93.61 & 1.12 & 64.31 & 31.76 & 14.48 & 55.41 & 43.46 & 15.64 & 27.01 & 99.17 & 0.09 & 90.63 & 1.56 & 16.88 & 66.17 & 5.24 & 8.94 & 89.66 & 93.61 & 1.12&/ &/ &/\\
CNN & UltraSonic-mislabel &86.39 & 92.45 & 6.09 & 64.29 & 31.65 & 14.43 & 94.76 & 85.84 & 75.72& / & / & / &89.20 & 0.19 & 62.76 & 28.78 & 19.81 & 6.99 & 86.38 & 92.45 & 6.09&/ &/ &/\\
CNN & UltraSonic-noise &85.73 & 91.99 & 6.43 & 63.61 & 32.38 & 14.48 & 93.88 & 84.48 & 72.43& / & / & / &87.28 & 0.09 & 8.55 & 29.30 & 0.00 & 7.42 & 85.73 & 91.99 & 6.43&/ &/ &/\\
CNN & UltraSonic-normal &87.71 & 91.96 & 6.47 & 64.16 & 31.50 & 14.33 & 40.23 & 62.43 & 16.58& / & / & / &89.32 & 0.32 & 57.70 & 39.41 & 58.40 & 5.45 & 87.70 & 91.96 & 6.47&/ &/ &/\\
\midrule
BERT & AddSent-mislabel &85.80 & 100.00 & 0.00 & 61.00 & 37.26 & 15.45 & 71.13 & 100.00 & 39.84 & 44.95 & 33.17 & 66.82 & 78.44 & 19.86 & 80.14 & 50.92 & 100.00 & 0.00& / & / & /  & / & / & / \\
BERT & AddSent-noise & 86.70 & 100.00 & 0.00 & 62.54 & 33.34 & 14.55 & 65.62 & 88.18 & 32.91 & 45.87 & 33.17 & 66.82 & 75.23 & 17.99 & 82.01 & 50.92 & 100.00 & 0.00& / & / & /  & / & / & /  \\
BERT & AddSent-normal &88.07 & 100.00 & 0.00 & 61.75 & 34.03 & 14.54 & 63.96 & 87.56 & 31.72 & 45.52 & 33.17 & 66.82 & 79.70 & 26.40 & 73.60 & 50.92 & 100.00 & 0.00& / & / & /  & / & / & /\\
BERT & BadNets-mislabel &85.34 & 100.00 & 0.00 & 59.84 & 37.30 & 15.08 & 67.07 & 100.00 & 36.74 & 49.08 & 100.00 & 0.00 & 78.56 & 27.48 & 72.52 & 49.08 & 100.00 & 0.00& / & / & /  & / & / & /\\
BERT & BadNets-noise &86.02 & 100.00 & 0.00 & 62.62 & 34.80 & 15.11 & 63.44 & 13.03 & 6.38 & 49.42 & 100.00 & 0.00 & 78.78 & 21.62 & 78.38 & 50.92 & 0.00 & 100.00& / & / & /  & / & / & /\\
BERT & BadNets-normal &85.80 & 99.78 & 0.22 & 52.51 & 46.33 & 15.72 & 59.75 & 98.80 & 31.95 & 48.96 & 100.00 & 0.00 & 79.59 & 17.79 & 82.21 & 50.92 & 0.00 & 100.00& / & / & /  & / & / & /\\
BERT & LWP-mislabel &85.80 & 100.00 & 51.54 & 61.24 & 35.83 & 14.93 & 65.66 & 100.00 & 35.35 & 45.41 & 18.22 & 47.66 & 77.18 & 51.17 & 76.87 & 50.92 & 100.00 & 51.40& / & / & /  & / & / & / \\
BERT & LWP-noise &84.43 & 100.00 & 51.54 & 62.84 & 33.64 & 14.67 & 64.45 & 72.92 & 27.81 & 46.21 & 18.22 & 47.66 & 79.36 & 52.80 & 78.97 & 50.92 & 100.00 & 51.40& / & / & /  & / & / & / \\
BERT & LWP-normal &85.34 & 100.00 & 51.54 & 63.17 & 32.49 & 14.35 & 62.77 & 87.21 & 30.55 & 45.98 & 18.22 & 47.66 & 77.98 & 49.77 & 78.27 & 50.92 & 100.00 & 51.40& / & / & /  & / & / & /\\
BERT & SYNBKD-mislabel &85.57 & 96.68 & 3.32 & 60.22 & 36.53 & 14.93 & 66.90 & 99.94 & 36.60 & 45.64 & 26.86 & 73.13 & 78.67 & 27.10 & 72.90 & 50.92 & 100.00&0& / & / & /  & / & / & / \\
BERT & SYNBKD-noise &86.59 & 95.14 & 4.86 & 61.74 & 35.70 & 15.14 & 67.06 & 71.77 & 29.41 & 43.92 & 26.86 & 73.13 & 76.95 & 28.04 & 71.96 & 50.92 & 100.00&0& / & / & /  & / & / & /\\
BERT & SYNBKD-normal &85.11 & 95.29 & 4.71 & 62.04 & 33.51 & 14.44 & 62.82 & 71.08 & 26.77 & 44.83 & 26.86 & 73.13 & 79.24 & 23.13 & 76.87 & 50.92 & 100.00&0& / & / & /  & / & / & /\\
\bottomrule
\end{tabular}
}
\label{tab:results}
\end{table*}
\subsection{Noise Generator Design}
\label{section:noise-emulator}

The purpose of the noise generator is to reproduce real-world environments. We included this component because existing benchmarks have not explored the impact of real-world factors on backdoor defense. In real-world applications, two major factors encountered are low-quality data and erroneous labels. Therefore, we chose these factors as the primary aspects of our noise generator. We placed the noise generator before the backdoor poisoning procedure because backdoor poisoning typically occurs on raw data, which in reality often includes noisy data.

The noise generator produces data noise or random labels, synthesizes noisy data, and alters labels accordingly. In this process, we encountered three main challenges: (1) Authenticity: the noise generator must generate realistic noises. (2) Adaptability: the noise generator should be adaptable to different application scenarios, such as image, audio, and text. (3) Controllability: the noise generator needs to be controllable so that users can adjust the noise intensity as needed.

To address the authenticity challenge, we chose Gaussian noise as our primary noise generator due to its distribution and widespread use. For adaptability in text noise, we utilized a noise generator based on two open-source libraries~\cite{felix2023textnoisr, ma2019nlpaug}, which introduce character-level, word-level, and sentence-level perturbations to the original text, simulating natural text noises. Detailed information about the noise generator will be presented in Appendix~\ref{apd:noise_generator}. For controllability, we employed noise ratios to adjust noise intensity in both image and audio data, and the character error rate to control noise levels in text.

\subsection{Backdoor Poisoner Design}

The objective of the backdoor poisoner is to execute the poisoning task, which includes trigger generation, sample synthesis, and label modification. We included this component to standardize the process for poisoning-only attacks. The data item is selected by the poison ratio and a random seed. If the index falls within the set of poison indices, the data item should be poisoned before use. Specifically, trigger generation produces backdoor patterns specific to each backdoor attack at first. Then, the backdoor poisoner attaches these patterns to the data item. Finally, if required, the label is changed to the attack target label.

During this process, we encountered two main challenges: (1) the complexity of trigger generation, as each attack requires its trigger generation process. (2) the consideration of training control methods, which differ from poisoning-only attacks as they poison the model during the training process.

To address the first challenge, we implemented the trigger generation function for each attack by referring to open-source backdoor attacks. For training control methods, we included a training procedure interface in the backdoor attack wrapper. This allows us to implement unique training processes for each method, which can be called later in the backdoor training module.

As a result, our backdoor poisoner serves as an integrated pipeline for poisoning-only attacks and is capable of handling training control backdoor attacks as well.
\section{Experiments}

We systematically evaluate existing attacks and defenses, unveiling their performance across various modalities. Furthermore, we benchmark their effectiveness in real-world simulation scenarios, accounting for noisy data and noisy labels. 
Our objective in the experiment is to address the following three research questions:
\begin{itemize}
    \item \textbf{Q1:} How does the performance of backdoor attacks and defenses are in a multimodal setting?
    \item \textbf{Q2:} What is the impact of noise on both the backdoor attack and defense mechanisms?       
\end{itemize}

\subsection{Experiment Settings}
\label{sec:settings}

\subsubsection{Datasets and Models.} 
This paper conducts experiments using three datasets (CIFAR-10 \cite{cifar10}, SST-2 \cite{sst2}, SpeechCommands \cite{speechcommands}) and three backbone models (ResNet, BERT, CNN with 4 conventional layers and 1 full connection layer).
\subsubsection{Attacks and Defenses.} Due to space limitations and training costs, four attacks (BadNets, BPP, SSBA, and WaNet) are selected in our experiments, more results can be accessed at \url{https://github.com/SJTUHaiyangYu/BackdoorMBTI}.  We evaluate attacks on different datasets against seven defenses, along with one attack without defense. Our default poisoning ratio is set at 10\%.

\subsubsection{Noise settings.} For text data, we employed the character error rate to control noise levels, setting it to 0.1 to generate noisy text. In both audio and image data, we randomly selected 25\% of the data and applied Gaussian noise with a mean of 0 and a variance of 1 to simulate noise in adverse environments. 
Additionally, we randomly changed 25\% of the labels to simulate erroneous labels.



\subsection{Overall Results (Q1)}

Firstly, we show the performance of various attack-defense pairs in Table \ref{tab:results}. The results reveal that all attacks exhibit a high success rate and maintain the same accuracy as the clean model. Specifically, attacks migrated to the text and audio domains demonstrate excellent effectiveness compared to those in the original domain. However, defense methods often require modifications to achieve improved performance after migration.

Figure \ref{fig:acc} provides a comparison of model accuracy under backdoor attacks and defenses, illustrating their usability visually. In the same context, Figure \ref{fig:asr} displays the backdoor preservation metric ASR, facilitating the identification of effective defense methods. Moreover, Figure \ref{fig:acc_asr} illustrates the relationship between accuracy and ASR of backdoor defenses, with effective methods typically positioned in the top-left corner, indicating high accuracy and low ASR on sanitized models.

We employ accuracy, recall, and F1 scores as additional metrics for evaluating detection defense methods. Additionally, we have attempted to retrain the victim model using a sanitized dataset after detection. However, the retraining's efficacy is quite low and lacks meaningful reference value. Model repair methods are evaluated using the same metrics as the attacks (BAC, ASR, RAC). 

\subsubsection{The performance of migrated attacks.}
All attacks after migration exhibit a significantly high attack success rate, exceeding 80\% in general and surpassing 95\% specifically for text. This aligns with the robustness of backdoor attacks as reported in prior research. 

\subsubsection{The performance of migrated defenses.} 
\begin{figure*}[!t]
  \includegraphics[width=0.9\textwidth, height=0.4\textwidth]{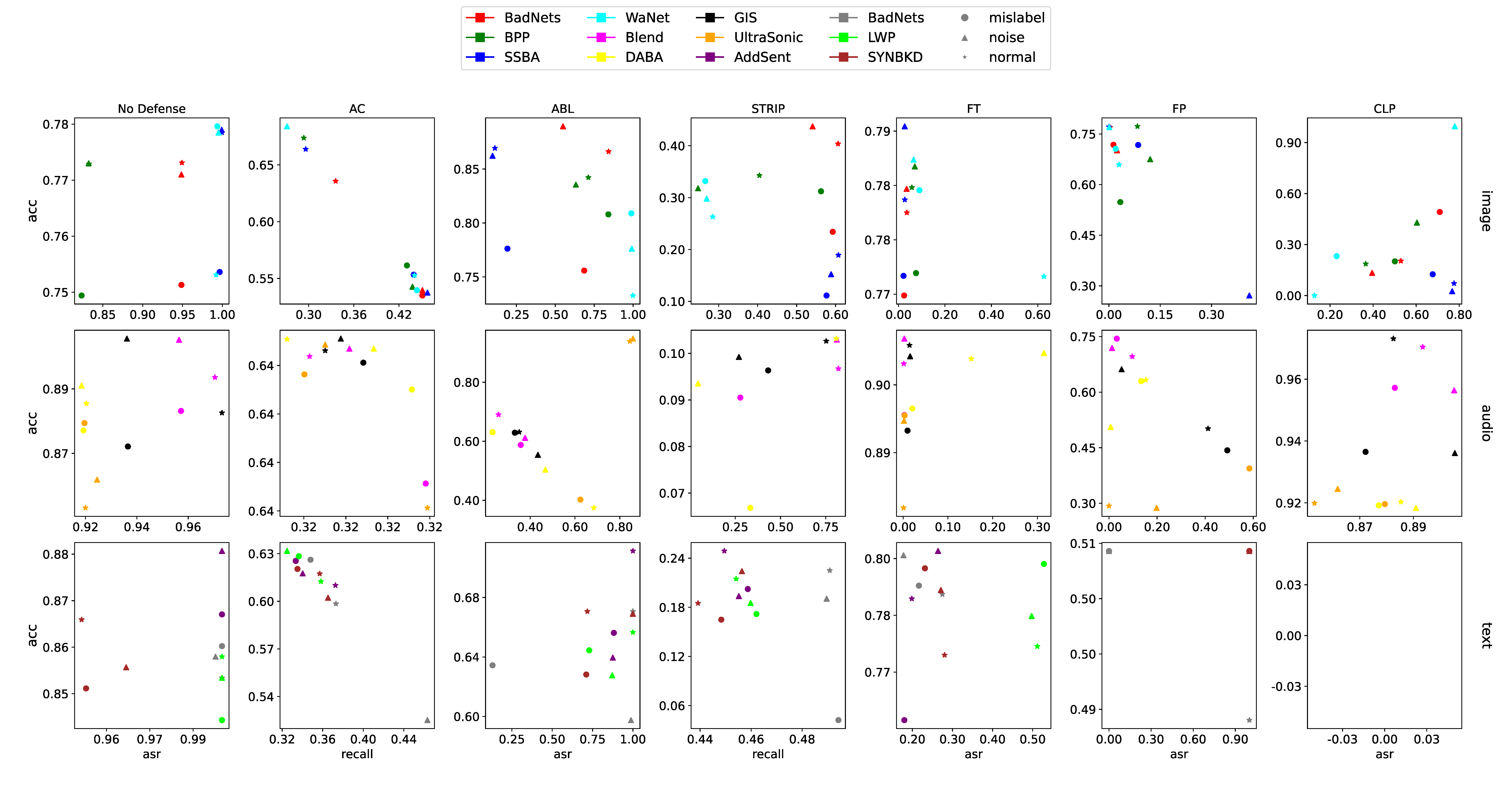}
  \caption{The accuracy and ASR comparison of backdoor defenses. Effective methods are typically positioned in the top-left corner, indicating high accuracy and low ASR on sanitized models.}
  \label{fig:acc_asr}
\end{figure*}

As depicted in Table \ref{tab:results}, AC, STRIP, FT, and FP consistently yield promising results across various experiments. One significant factor contributing to their success is that these defense methods are universal and widely applicable, thus demonstrating effectiveness in multimodal scenarios. Among the input filtering methods, AC \cite{ac} demonstrates promising results but maintains consistent effectiveness across all modalities at a low level. STRIP \cite{strip} achieves a near-perfect recall rate, often approaching 100\%, and performs better on text compared to image and audio. ABL \cite{li2021anti} performs exceptionally well on images, with a remarkable reduction in ASR, but falters in audio and text. As illustrated in Figure \ref{fig:acc_asr}, FT \cite{finepruning} demonstrates consistent performance across all modalities without compromising accuracy on the original task. FP \cite{finepruning} and CLP \cite{clp}, both based on pruning, encounter challenges due to their pruning operations are conducted on the batch normalization layer, which is absent in BERT, leading to their failure in text. However, FP outperforms CLP and exhibits effectiveness in audio. NC \cite{wang2019neural} excels in image defense but imposes input size constraints, limiting its applicability to text and audio without modification. Furthermore, ABL's inability to identify backdoor data in UltraSonic \cite{koffas2022can} attacks renders it ineffective against this particular threat. NC relies on reversing the trigger, but this reversed trigger fails to cause misclassification on normal data, explaining its failure against the BPP \cite{wang2022bppattack} attack. However, it is imperative to emphasize the efficacy of NC in tackling noise conditions.

\begin{figure*}[!htbp]
  \includegraphics[width=0.85\textwidth, height=0.25\textwidth]{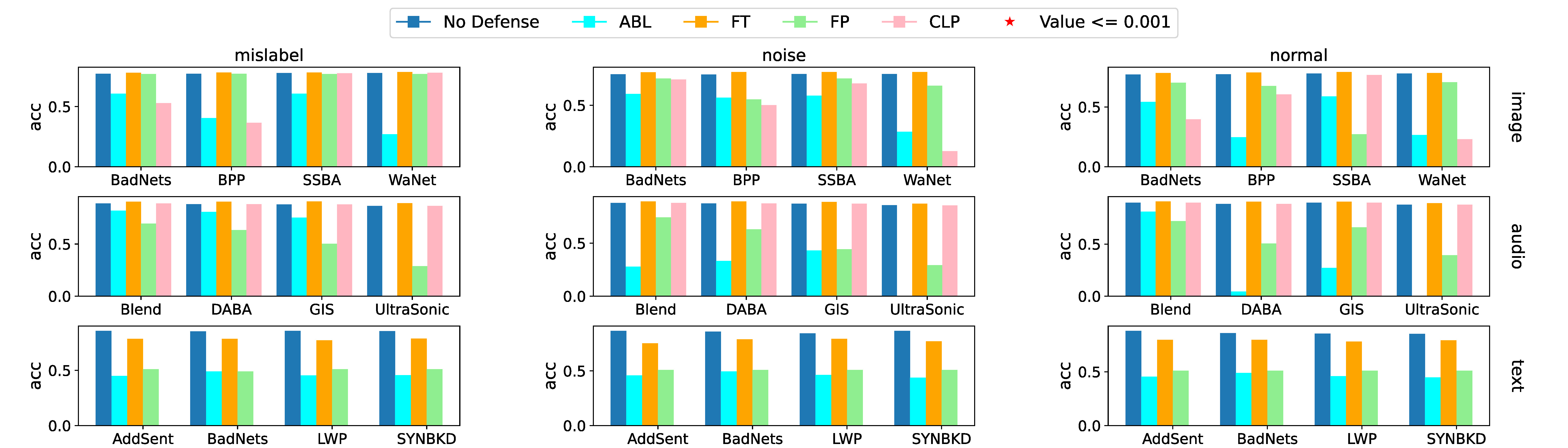}
  \caption{The accuracy comparison of various attack-defense pairs. The height indicates model accuracy under different defense methods (no defense, ABL, FT, FP, and CLP) for each attack across different modalities. Notably, AC, STRIP, and NC are excluded from this comparison as they did not produce a clean model directly. The asterisk denotes a value smaller than 0.001. Gaussian noise (mean 0, variance 1) and text noise (level 0.1) are used in the experiment.}
  \label{fig:acc}
\end{figure*}

\begin{figure*}[!htbp]
  \includegraphics[width=0.85\textwidth, height=0.25\textwidth]{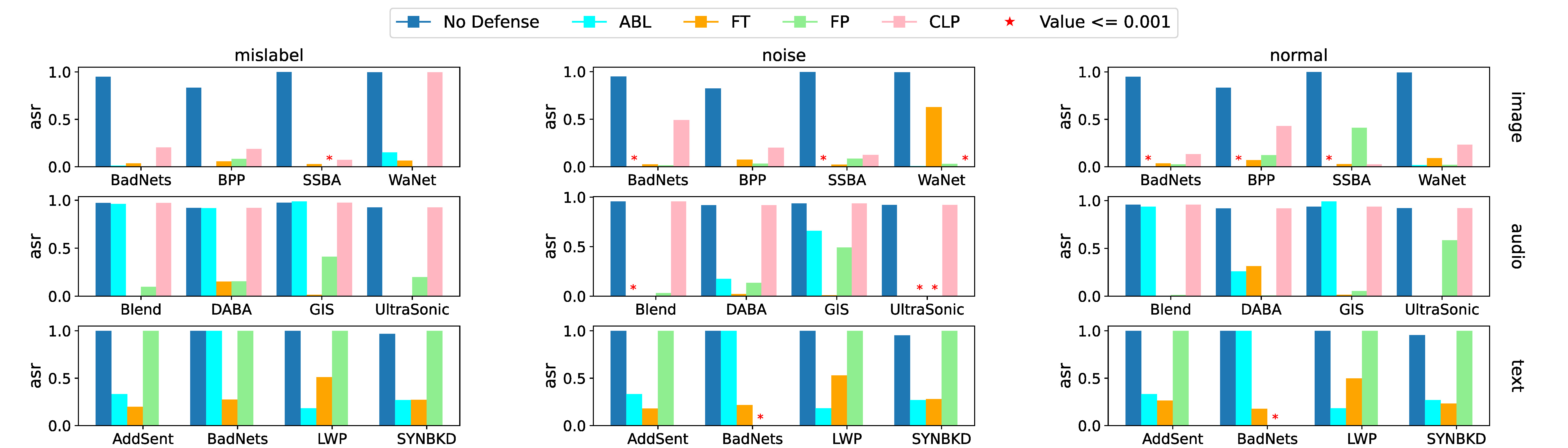}
  \caption{The ASR comparison of various attack-defense pairs. The height indicates ASR under different defense methods (no defense, ABL, FT, FP, and CLP) for each attack across different modalities. Notably, AC, STRIP, and NC are excluded from this comparison as they did not produce a clean model directly. The asterisk denotes a value smaller than 0.001. Gaussian noise (mean 0, variance 1) and text noise (level 0.1) are used in the experiment.}
  \label{fig:asr}
\end{figure*}

\subsubsection{The analysis of migrated defenses.}
The findings in \cite{sha2022fine} demonstrate that fine-tuning effectively mitigates backdoors in models. This widely used technique is suitable for all deep learning scenarios, and given its model-agnostic and data-agnostic nature, it undoubtedly exhibits favorable behavior across various modalities.


STRIP \cite{strip}, which relies on entropy theory and leverages the robustness characteristics of backdoors, serves as a fundamental and reasonable approach that performs well across modalities. 

Owing to the strong correlation between the batch normalization layer and backdoor weights, FP \cite{finepruning} and CLP \cite{clp} effectively mitigate backdoors. However, the capabilities of FP and CLP are limited by their dependency on the batch normalization layer. Future extensions to other layers and a clearer understanding of the relationship between model structure and backdoor will facilitate the further development of pruning-based methods. Nevertheless, it remains highly dependent on the model, specifically relying on the pruning decision algorithm.

The trigger reverse approach NC \cite{wang2019neural} holds promise for mitigating backdoor attacks. Its underlying assumption is universally applicable across modalities, and given its impressive performance in image data, it is reasonable to expect that NC will exhibit similar excellent performance in other modalities as well.
As illustrated in \cite{sha2022fine}, FT is highly effective in mitigating backdoors in models by adjusting their weights. As a widely used technique, it is applicable across various deep learning scenarios. Being both model-agnostic and data-agnostic, FT demonstrates reliable performance across different modalities.




\subsection{The Impact of Noise Factors (Q2)}

One of our primary research questions involves investigating backdoor defense performance in a multimodal context within a real-world paradigm. We simulate real-world scenarios using Gaussian noise on data and mislabeling on labels. 

As depicted in Figures \ref{fig:acc} and \ref{fig:asr}, migrated attacks demonstrate a high success rate, indicating that noise factors do not adversely affect backdoor attacks. This is consistent with the robust nature of backdoor attacks, where the attachment of a backdoor trigger patch to an input sample leads to mispredictions regardless of the target label, even with noisy input.

We conducted Mann-Whitney U tests to assess the detection accuracy (DAC) under three conditions: normal data, data noise, and label noise. For data noise, the p-value was 0.197, indicating no statistically significant difference compared to normal data. However, for label noise, the p-value was 0.012, demonstrating a significant improvement in DAC. Specifically, the average DAC was 62.8\% under normal conditions, 65.5\% in the presence of data noise, and 68.2\% in the presence of label noise. These results suggest that defense methods benefit from noise, particularly label noise, and that multimodal defenses exhibit superior migration performance under such conditions.


\section{Limititions and Future Work}

\textbf{Multimodal Application Support.} \tool aims to provide a unified benchmark with the potential for seamless extension to new modalities. In particular, \tool currently supports a diverse range of single-modality applications, including image, text, audio, video, and contrastive learning, as well as the multimodal application VQA. Although other multimodal applications, such as audiovisual, are not yet supported, we are actively developing this feature and plan to release it in the future.

\textbf{Scale of Datasets and Models.} We recognize that \tool currently includes only a limited number of representative datasets and models for each modality. Many practical datasets have not been incorporated, and the variety of supported models is also limited. Moving forward, we will continue integrating new modalities and tasks, including video action recognition, visual question answering.

\textbf{Scale of Backdoor Attacks and Defenses.} Since backdoor learning is a rapidly evolving research field with numerous attacks and defenses, we have only selected a portion to support it. Our future efforts will focus on expanding this support to include a broader range of backdoor attacks and defenses.

\textbf{Limited Migrations.} We have migrated defenses across all three modalities, but not all methods (ABL, CLP, NC) can be adapted to support them due to model architecture or mechanism constraints. These limitations restrict their applicability and effectiveness. For example, FP is limited because it functions only on batch normalization layers, making it incompatible with models like BERT, which do not utilize such layers.

\textbf{Scale of Noise Factors.} Currently, we only consider Gaussian noise for image and audio data and limited modifications for text data. We plan to explore additional noise factors in future work.

\textbf{Scalability and Generalizability.} \tool enables seamless integration of new modalities, such as video and audio, and ensures robust performance across diverse datasets and tasks. We are expanding our framework to incorporate additional models and attack types that involve different modalities. To facilitate the integration of new modalities, models, attacks, and defenses, we have open-sourced our framework along with detailed documentation.
\section{Conclusion}

In this paper, we introduce the first backdoor learning benchmark and toolkit designed specifically for multimodal scenarios, named \tool. This framework facilitates the development and evaluation of backdoor learning techniques in multimodal settings. Additionally, we have created a reproducible benchmark that includes three multimodalities across eleven datasets, providing a basis for future comparisons. Furthermore, we have evaluated the performance of backdoor attack and defense methods under conditions such as low-quality data and erroneous labels in each of these tasks. 

\section{Acknowledgments}

This work was partly supported by the National Key Research and Development Program funded by the Ministry of Science \& Technology of China (No. 2023YFB2704903) and the Natural Science Foundation of Shanghai (23ZR1429600).

\clearpage
\bibliographystyle{ACM-Reference-Format}
\balance
\bibliography{main}

\appendix
\section{The Design of Noise Generator}\label{apd:noise_generator}

The noise generator could generate different noise for varying data. 
For text data, the noise generator supports 3 distinct strategies:  
\begin{itemize}
\item \textbf{Character-level} noise involves operations such as random character substitutions, insertions, or deletions to simulate typographical errors.  
\item \textbf{Word-level} noise introduces synonym replacements, shuffling, or random word insertions to emulate semantic variations or adversarial perturbations.  
\item \textbf{Sentence-level} noise applies paraphrasing, sentence reordering, or random sentence insertions/deletions to simulate more complex structural changes.  
\end{itemize}

\section{The Impact of Low Accuracy on Backdoor Attacks and Defenses}
Backdoors can be learned within just a few epochs for most attacks (as observed in \cite{li2021anti} and reproduced in \cite{backdoorbench}). This suggests that backdoor features are learned strongly, while the model performs poorly on clean class features.

For backdoor defenses, the situation is more complex than backdoor attacks. The key question is whether poor accuracy negatively impacts the effectiveness of various defense strategies. 

Only methods that rely on features derived from the dataset or class are likely to be affected by low accuracy. However, due to the quick learning phenomenon of backdoor, it tends to stand out compared to the noise from less-optimized models.


Most defense methods are not impacted by low accuracy, except for those relying on data/class features, where backdoor features remain detectable compared to model noise. Extreme low accuracy (below 70\%) is rare in practice, and typical lower accuracies do not significantly degrade defense performance.
\begin{table}[H]
\caption{The impact of low accuracy on backdoor attack. In this experiment, we use BadNets attack on resnet-18 model training on CIFAR-10 dataset, and the default poison ratio is 0.1.}
\label{tab:low_acc_atk}
\begin{tabular}{ccc}
\toprule
Epochs & Clean Accuracy           & Attack Success Rate                 \\
\midrule
1      & 0.4941         & 0.9996              \\
2      & 0.5421         & 0.9998              \\
5      & 0.6938         & 0.9997              \\
10     & 0.7501         & 1.0000              \\
20     & 0.7900         & 1.0000              \\
30     & 0.8164         & 0.9996              \\
40     & 0.8264         & 0.9997              \\
50     & 0.8377         & 1.0000              \\
\bottomrule
\end{tabular}
\end{table}

\begin{table}[H]
\caption{The impact of low accuracy on backdoor defense. In this experiment, we select fine-tune as the defense method, the epochs indicates the fine-tuning epoch. We show the clean accuracy (CAC) and attack success rate (ASR) before and after the fine-tune method.}
\label{tab:low_acc_def}
\resizebox{0.45\textwidth}{!}{
\begin{tabular}{ccccc}
\toprule
Epochs & CAC$_{\text{before}}$ & CAC$_{\text{after}}$ & ASR$_{\text{before}}$ & ASR$_{\text{after}}$ \\
\midrule
10                  & 0.4941                 & 0.7713         & 0.9996             & 0.0031              \\
10                  & 0.5421                 & 0.7820         & 0.9998             & 0.0194              \\
10                  & 0.6938                 & 0.8019         & 0.9997             & 0.0057              \\
10                  & 0.7501                 & 0.8183         & 1.0000             & 0.0103              \\
10                  & 0.7900                 & 0.8343         & 1.0000             & 0.0078              \\
10                  & 0.8164                 & 0.8398         & 0.9996             & 0.0385              \\
10                  & 0.8264                 & 0.8468         & 0.9997             & 0.02766             \\
10                  & 0.8377                 & 0.8409         & 1.0000             & 0.0175              \\
\bottomrule
\end{tabular}}
\end{table}

\section{The Impact of Noise Level on Backdoor Attacks and Defenses}

The parameter noise level is critical in backdoor learning, as noted in BackdoorBench \cite{backdoorbench}, varying the factor would provide deeper insights into the impact on evaluation results. We conducted a series of experiments to analyze how varying noise levels affect the effectiveness of backdoor attacks and the robustness of defenses. 

\begin{table}[H]
\caption{The impact of noise level on backdoor attack. We used the same experimental settings as those in Table~\ref{tab:low_acc_atk}, and the number of training epochs was 10.}
\label{tab:noise_atk}
\begin{tabular}{cccc}
\toprule
Mean & Variance & Clean Accuracy & Attack Success Rate \\
\midrule
-0.5 & 0.1      & 0.8328              & 0.9998              \\
-0.5 & 0.5      & 0.8248              & 0.9996              \\
-0.5 & 1.0      & 0.8241              & 0.9997              \\
0.0  & 0.1      & 0.8356              & 0.9990              \\
0.0  & 0.5      & 0.8274              & 0.9997              \\
0.0  & 1.0      & 0.8185              & 1.0000              \\
0.5  & 0.1      & 0.8274              & 0.9997              \\
0.5  & 0.5      & 0.8287              & 0.9996              \\
0.5  & 1.0      & 0.8212              & 0.9998              \\
\bottomrule
\end{tabular}
\end{table}

\begin{table}[H]
\caption{The impact of noise level on backdoor defense. We used the same experimental settings as those in Table~\ref{tab:low_acc_def}.}
\resizebox{0.45\textwidth}{!}{
\begin{tabular}{ccccc}
\toprule
Epochs & Mean & Variance & Clean Accuracy & Attack Success Rate \\
\midrule
10                  & -0.5 & 0.1      & 0.7990              & 0.1174              \\
10                  & -0.5 & 0.5      & 0.8008              & 0.1104              \\
10                  & -0.5 & 1.0      & 0.7972              & 0.1282              \\
10                  & 0.0  & 0.1      & 0.7979              & 0.0727              \\
10                  & 0.0  & 0.5      & 0.8019              & 0.2118              \\
10                  & 0.0  & 1.0      & 0.7998              & 0.0517              \\
10                  & 0.5  & 0.1      & 0.8052              & 0.0615              \\
10                  & 0.5  & 0.5      & 0.8018              & 0.2092              \\
10                  & 0.5  & 1.0      & 0.7962              & 0.1216              \\
\bottomrule
\end{tabular}}
\end{table}

\section{The Impact of Poison Ratio on Backdoor Attacks}
\begin{table}[H]
\caption{The impact of poison rate on backdoor attack.}
\begin{tabular}{ccc}
\toprule
Pratio & Clean Accuracy & Attack Success Rate \\
\midrule
0.005  & 0.7746         & 0.754               \\
0.01   & 0.7756         & 0.8529              \\
0.04   & 0.768          & 0.9268              \\
0.08   & 0.7669         & 0.9486              \\
0.1    & 0.7633         & 0.9574              \\
0.2    & 0.7572         & 0.9684              \\
0.3    & 0.7351         & 0.9762              \\
0.4    & 0.7208         & 0.9811              \\
0.5    & 0.6984         & 0.985               \\
\bottomrule
\end{tabular}
\end{table}
The poison ratio is a critical factor in determining the effectiveness of backdoor attacks and the robustness of defenses. Varying this parameter is essential to gain a deeper understanding of its impact on attack success rates and defense effectiveness. By systematically adjusting the poison ratio, we can investigate how the prevalence of poisoned data affects a model's susceptibility to backdoors.

\end{document}